\documentclass[10pt,aps,prb,twocolumn]{revtex4-2}
\usepackage[english]{babel}
\usepackage[T1]{fontenc} 
\usepackage[utf8]{inputenc} 
\usepackage{verbatim}
\usepackage{graphicx}
\usepackage{graphics}
\usepackage{color}
\usepackage{textcomp}
\usepackage{amsfonts}
\usepackage{amsmath}
\usepackage{hyperref}
\usepackage{amssymb}
\usepackage{mathtools}
\usepackage{tabularx}
\usepackage[a4paper, margin=1.5cm]{geometry}

\newcolumntype{Y}{>{\centering\arraybackslash}X}

\begin{document}

\title{Generalized Diffusive Epidemic Process with Permanent Immunity in Two Dimensions}

\author{V. R. Carvalho}
\affiliation{Departamento de F\'{\i}sica, Universidade Federal do Piau\'{i}, 57072-970, Teresina - PI, Brazil}
\author{T. F. A. Alves}
\affiliation{Departamento de F\'{\i}sica, Universidade Federal do Piau\'{i}, 57072-970, Teresina - PI, Brazil}
\author{G. A. Alves}
\affiliation{Departamento de F\'{i}sica, Universidade Estadual do Piau\'{i}, 64002-150, Teresina - PI, Brazil}
\author{D. S. M. Alencar}
\affiliation{Departamento de F\'{\i}sica, Universidade Federal do Piau\'{i}, 57072-970, Teresina - PI, Brazil}
\author{F. W. S. Lima}
\affiliation{Departamento de F\'{\i}sica, Universidade Federal do Piau\'{i}, 57072-970, Teresina - PI, Brazil}
\author{A. Macedo-Filho}
\affiliation{Departamento de F\'{i}sica, Universidade Estadual do Piau\'{i}, 64002-150, Teresina - PI, Brazil}
\author{R. S. Ferreira}
\affiliation{Departamento de Ci\^{e}ncias Exatas e Aplicadas, Universidade Federal de Ouro Preto, 35931-008, Jo\~{a}o Monlevade - MG, Brazil}

\date{Received: date / Revised version: date}

\begin{abstract}

We introduce the generalized diffusive epidemic process, which is a metapopulation model for an epidemic outbreak where a non-sedentary population of walkers can jump along lattice edges with diffusion rates $D_S$ or $D_I$ if they are susceptible or infected, respectively, and recovered individuals possess permanent immunity. Individuals can be contaminated with rate $\mu_c$ if they share the same lattice node with an infected individual and recover with rate $\mu_r$, being removed from the dynamics. Therefore, the model does not have the conservation of the active particles composed of susceptible and infected individuals. The reaction-diffusion dynamics are separated into two stages: (i) Brownian diffusion, where the particles can jump to neighboring nodes, and (ii) contamination and recovery reactions. The dynamics are mapped into a growing process by activating lattice nodes with successful contaminations where activated nodes are interpreted as infection sources. In all simulations, the epidemic starts with one infected individual in a lattice filled with susceptibles. Our results indicate a phase transition in the dynamic percolation universality class controlled by the population size, irrespective of diffusion rates $D_S$ and $D_I$ and a subexponential growth of the epidemics in the percolation threshold.

\end{abstract}



\maketitle

\section{Introduction}

This work considers the diffusive epidemic process (DEP) with permanent immunity. The standard DEP has the feature of a metapopulation model\ \cite{colizza_reactiondiffusion_2007}, which the lattice is a substrate to a fluctuating infected population\ \cite{van_wijland_wilson_1998, fulco_critical_2001, bertrand_critical_2007, maia_diffusive_2007, dickman_nature_2008, da_silva_critical_2013-2, tarpin_nonperturbative_2017, argolo_stationary_2019, alves_diffusive_2021, alencar_two-dimensional_2023}, unlike sedentary models where the individuals are the lattice nodes on typical stochastic lattice processes\ \cite{tome_stochastic_2018}. In the standard DEP, individuals can diffuse by hopping to neighboring sites and interact only when they divide the same lattice node. The DEP on lattices has a rich critical behavior, whose exponents should differ according to three regimes: $D_S>D_I$, $D_S=D_I$, and $D_S<D_I$ where $D_S$ and $D_I$ are the diffusion rates of the susceptible and infected individuals, respectively\ \cite{kree_effects_1989,van_wijland_wilson_1998, fulco_critical_2001, tarpin_nonperturbative_2017, alencar_two-dimensional_2023}. The critical behavior of standard DEP in any diffusion regime also differs from the paradigmatic directed percolation universality class\ \cite{henkel_non-equilibrium_2008, tarpin_nonperturbative_2017}.

In general, one can simulate the DEP as a cellular automata model where reaction-diffusion dynamics can be separated into two synchronous stages\ \cite{fulco_critical_2001, da_silva_critical_2013-2, alencar_two-dimensional_2023}. In diffusion, susceptible and infected individuals jump to neighboring nodes according to the $D_S$ and $D_I$ rates, respectively. All lattice nodes should update their populations after the diffusion stage. Eventually, the reaction stage takes place according to the channels
\begin{equation}
   \left\lbrace
   \begin{aligned}
      &S + I \stackrel{\mu_c}{\longrightarrow} 2I, \\
      &I \stackrel{\mu_r}{\longrightarrow} R, \\
   \end{aligned}
   \right.
   \label{sir-channels}
\end{equation}
where Eq.\ (\ref{sir-channels}) means that susceptible individuals would be contaminated if they share the same lattice node with at least one infected individual with rate $\mu_c$ and infected individuals can spontaneously recover with rate $\mu_r$. According to the standard definition of DEP\ \cite{fulco_critical_2001}, individuals become susceptible before they recover (compartment $R$ is just the same as $S$), which means the susceptible individuals can be infected again. We can turn recovered individuals in compartment $R$ permanently immune to the disease to prevent them from interacting with the active particles of compartments $S$ and $I$. Therefore, the population of active particles will not be conserved after the modification.

In sedentary stochastic epidemic models, perfect immunity is introduced in the generalized epidemic process (GEP)\ \cite{mollison_spatial_1977, grassberger_spreading_1997}. The GEP model describes the spreading of a non-conserved population and falls into the dynamic percolation universality class\ \cite{j_l_cardy_epidemic_1985, grassberger_spreading_1997, henkel_non-equilibrium_2008}. Permanent immunity generally makes a stochastic process fall into the dynamic percolation universality class\ \cite{janssen_field_2005}. Permanent immunity is also essential to introduce the effects of vaccination protocols\ \cite{wang_statistical_2016}. Recently, some reported results of a vaccination epidemic model using Monte Carlo (MC) simulations, finite-size scaling (FSS), and mean-field approximations\ \cite{pires_dynamics_2017} presented continuous phase transitions for square lattices with critical exponents compatible with the directed percolation universality class. 

An interesting question is the role of permanent immunity when added to the general DEP dynamics and metapopulation models. Therefore, we analyze the effect of permanent immunity in the DEP. We define the analogous model for diffusive individuals, called the generalized diffusive epidemic process (GDEP), and study its properties. By activating lattice nodes with successful contaminations, we address the problem of mapping the epidemic dynamics into a growing process and exploring it as a percolation problem. In addition, the final removed population and typical related observables are analyzed. 

The definition of GDEP presented in this manuscript is close to one extension of the susceptible-infected-removed (SIR) model\ \cite{paoluzzi_single-agent_2021} where mobile point particles are free to move in a square space with periodic boundary conditions and interact according to the reaction channels in Eq.\ (\ref{sir-channels}) if they share a single square cell, a subdivision of the square space. The main difference is the diffusive movement of the individuals is restricted to a two-dimensional lattice in the GDEP, like in transportation networks\ \cite{newman_networks_2018}. Limiting particle movement to a lattice allows one to see the dependence of the critical exponents on the lattice dimension\ \cite{stanley_introduction_1987, amit_field_2005} and also analyze the phase transition as controlled by the population size. One can expect the percolation transition to happen when one increases the particle concentration.

We organized this paper as follows. In Sec. \ref{sec-model} is presented the Markovian dynamics of the GDEP model. In Sec. \ref{sec-results} are discussed our numerical results. Finally, we give our final considerations in Sec. \ref{sec-conclusions}.

\section{Model and Scaling}\label{sec-model}

This section presents the GDEP at a square lattice with $N=L^2$ nodes. The reaction-diffusion dynamics is formulated as a cellular automata model and separated into two synchronous stages where all lattice nodes are updated simultaneously. The dynamics are summarized in the following kinetic Monte-Carlo rules\ \cite{dorogovtsev_critical_2008, pastor-satorras_epidemic_2015}:
\begin{enumerate}
   \item We begin by placing the walker population in the lattice nodes. The walker population size $N_w$ is related to the concentration $\rho$ as
      \begin{equation}
         N_{w} = \rho N,
      \end{equation}
   Increasing the concentration $\rho$, a continuous phase transition from a non-percolating phase to a percolating one will happen, in a critical threshold $\rho_c$. Dynamics start from a state with one infected individual and $N_w-1$ susceptibles and the stochastic arrays $\textbf{S} = \lbrace S_1, S_2, ... , S_N \rbrace$ and $\textbf{I} = \lbrace I_1, I_2, ... , I_N \rbrace$ store the susceptible and infected populations per node, respectively;
   \item In a single step, the arrays $\textbf{S}$ and $\textbf{I}$ are updated at the end of each stage:
   \begin{itemize}
     \item \textbf{Diffusion:} susceptible and infected individuals in all nodes jump with probabilities $D_S$ and $D_I$, respectively, along one lattice edge;
     \item \textbf{Reactions}: if susceptible individuals share a lattice node with at least one infected individual, they will be infected at a rate $\mu_c$. In addition, infected sites are removed with probability $\mu_r$. In all cases, we used $\mu_c=\mu_r=0.5$;
     \end{itemize}
   \item Step 2 is iterated until the system has no infected individual. The system will always reach an absorbing state with a lattice filled only with removed and susceptible individuals.
\end{enumerate}
The Monte Carlo time unit is the time-lapse of updating all lattice nodes.

The Markovian rules above present the following mean-field behavior\ \cite{fulco_critical_2001}
\begin{equation}
   \left\lbrace 
   \begin{aligned}
      &\frac{\partial \rho_S}{\partial t} = D_S \nabla^2 \rho_S - \mu_c \rho_S \rho_I, \\
      &\frac{\partial \rho_I}{\partial t} = D_I \nabla^2 \rho_I + \mu_c \rho_S \rho_I - \mu_r \rho_I, \\
      &\frac{\partial \rho_R}{\partial t} = D_R \nabla^2 \rho_R + \mu_r \rho_I, \\
   \end{aligned}
   \right.
\end{equation}
where $\rho_{\text{S,I,R}}$ are the particle densities in each compartment $S$, $I$, and $R$, which obey the following conservation law
\begin{equation}
   \rho_S+\rho_I+\rho_R = \rho.
\end{equation}
One can see the striking similarity with the mean-field equations for the susceptible-infected-removed (SIR) model\ \cite{de_souza_new_2011, tome_critical_2010, santos_epidemic_2020, alencar_epidemic_2020}, which motivates us to map the model into a growing model where we can use union-find algorithms to study the system as a percolation problem. The mapping is done by activating nodes with successful contaminations in a GDEP realization. One of these union-find algorithms is the Newman-Ziff algorithm\ \cite{newman_efficient_2000, newman_fast_2001, yang_alternative_2012}, which is suitable for finding the cluster-size distribution $n_{\mathrm{cluster}}(s)$.  In the following, we call the nodes with at least one successful contamination as activated ones. 

We can interpret the activated nodes as infection sources in a lattice of connected places. Given the Brownian nature of the epidemic spreading by the random walkers, the activity is concentrated at the cluster boundaries. We also expect only one giant component in every realization of GDEP, which comprises all the epidemic spreading. The order parameter of the percolation transition is the wrapping probability\ \cite{de_souza_new_2011, tome_critical_2010, santos_epidemic_2020, alencar_epidemic_2020}
\begin{equation}
   P = \left< P_\infty \right>.
   \label{orderparameter}
\end{equation}
The $\left< \right>$ indicates an average over repetitions of GDEP dynamics starting from only one seed. The averaged order parameter is the wrapping probability of a cluster formed by the activated nodes.

From the cluster size distribution $n_{\mathrm{cluster}}(s)$, we can also obtain the cluster size moments which are given by
\begin{equation}
   \left< s^\ell \right> = \frac{1}{N_r}\sum_s s^{\ell+1} n_{\mathrm{cluster}}(s,\rho),
   \label{cluster-moments}
\end{equation}
where $N_r$ is the number of activated nodes. Taking the sum on Eq.\ (\ref{cluster-moments}), excluding the wrapping cluster to calculate the first moment, we have the mean cluster size or the susceptibility\ \cite{santos_epidemic_2020, alencar_epidemic_2020}
\begin{equation}
   \chi = \left< s \right>.
   \label{susceptibility}
\end{equation}
Now, taking the sum including the wrapping cluster\ \cite{de_souza_new_2011}, we can define a ratio analogous to Binder cumulants for spin models
\begin{equation}
   U = \left< P_\infty \right> \frac{ \left< s^2 \right>}{ \left< s \right> ^2}.
   \label{bindercumulant}
\end{equation}

The cluster moments present the following finite-size behavior\ \cite{stauffer_introduction_2018, christensen_complexity_2005}
\begin{equation}
   \left< s^\ell \right> \approx L^{(\ell-1)\beta/\nu+\ell\gamma/\nu}f_{s_\ell}\left( L^{1/\nu} \left| \rho - \rho_c \right| \right),
   \label{cluster-moments-scaling}
\end{equation}
from we can obtain the following finite-size relations
\begin{equation}
   \begin{aligned}
      &U    \approx f_{U}\left(L^{1/\nu}\left\vert \rho - \rho_{c} \right\vert\right),                  \\
      &P    \approx L^{-\beta/\nu}f_{P}\left(L^{1/\nu}\left\vert \rho - \rho_{c} \right\vert\right),    \\
      &\chi \approx L^{\gamma/\nu}f_{\chi}\left(L^{1/\nu}\left\vert \rho - \rho_{c} \right\vert\right), \\
   \end{aligned}
   \label{cluster-scaling}
\end{equation}
where $f_{U,P,\chi}$ are scaling functions.

Also, one can collect the average removed population and its moments to investigate the critical behavior of GDEP, which should present the same critical behavior as the cluster observables. As already stated, the GDEP dynamics form only one giant component because of the Brownian diffusion of the individuals. Therefore, the removed population can be interpreted as the cluster mass in a way that removed population moments should obey the following finite-size scaling relation
\begin{equation}
   \left< R^\ell \right> \approx L^{(\ell-1)\beta/\nu+\ell\gamma/\nu}f_{R_\ell}\left( L^{1/\nu} \left| \rho - \rho_c \right| \right).
   \label{removed-moments-scaling}
\end{equation}
We can define the following observables
\begin{equation}
   \begin{aligned}
      &U^\prime = \frac{\left< R^2 \right>^2}{\left< R \right> \left< R^3 \right>}, \\
      &r        = \frac{\left< R \right>}{N_w},                                     \\
      &\psi     = \frac{\left< R^2 \right>}{\left< R \right>^2},                    \\
   \end{aligned}
   \label{observables-infection}
\end{equation}
In Eq.\ (\ref{observables-infection}), $r$ is the average removed concentration. From Eq.\ (\ref{removed-moments-scaling}), we have the following finite-size behavior of the observables in Eq.\ (\ref{observables-infection})
\begin{equation}
   \begin{aligned}
      &U^\prime \approx f_{U^\prime}\left(L^{1/\nu}\left\vert \rho - \rho_{c} \right\vert\right),          \\
      &r        \approx L^{\gamma/\nu-2}f_{r}\left(L^{1/\nu}\left\vert \rho - \rho_{c} \right\vert\right), \\
      &\psi     \approx L^{\beta/\nu}f_{\psi}\left(L^{1/\nu}\left\vert \rho - \rho_{c} \right\vert\right). \\
   \end{aligned}
   \label{removed-compartment-scaling}
\end{equation}

Moreover, we should have a dynamical critical behavior associated with dynamical exponents. One can investigate dynamical exponents and the dynamical behavior by the following set of observables of GDEP dynamics\ \cite{wada_critical_2015}, which scales as
\begin{equation}
   \begin{aligned}
      &P_s(t) \sim t^{-\delta}, \\
      &r(t) \sim t^{\eta},      \\
   \end{aligned}
   \label{dynamical-properties}
\end{equation}
at the critical threshold $\rho_c$ where $P_s(t)$ is the survival probability, i.e., the fraction of GDEP realizations still active after a Monte-Carlo time. The second one is the total number of removed individuals as a function of the time. The survival probability decays as a power-law at the critical threshold $\rho_c$. In addition, the number of infected individuals increases as a power-law at the critical threshold $\rho_c$. 

On finite lattices, one can expect the following finite-size critical behavior of the dynamical properties in Eq.\ (\ref{dynamical-properties}
\begin{equation}
   \begin{aligned}
      &P_s(t) \approx L^{-\beta/\nu} g_{P_s}\left( L^{-z} t \right), \\
      &r(t) \approx L^{\gamma/\nu-2} g_{r}\left( L^{-z} t \right),  \\
   \end{aligned}
   \label{dynamical-scaling}
\end{equation}
where $g_{P_s,r}$ are scaling functions. The dynamical critical exponents $z$, $\delta$, and $\eta$ in Eqs.\ (\ref{dynamical-properties}) and (\ref{dynamical-scaling}) are related with 2d percolation exponents by the scaling relations $\delta z = \beta/\nu$ and $\eta z = \gamma/\nu$.

\section{Results and Discussion}\label{sec-results}

We performed Monte Carlo simulations of the GDEP on square lattices. In all realizations, we begin from one seed. The dynamics would end when the system falls into one of the possible absorbing states, with no remaining infected individual. For each seed, the GDEP grows one cluster formed by the activated nodes. All curves are given as functions of node concentration $\rho$.

\begin{figure*}[ht!]
   \includegraphics[scale=0.35]{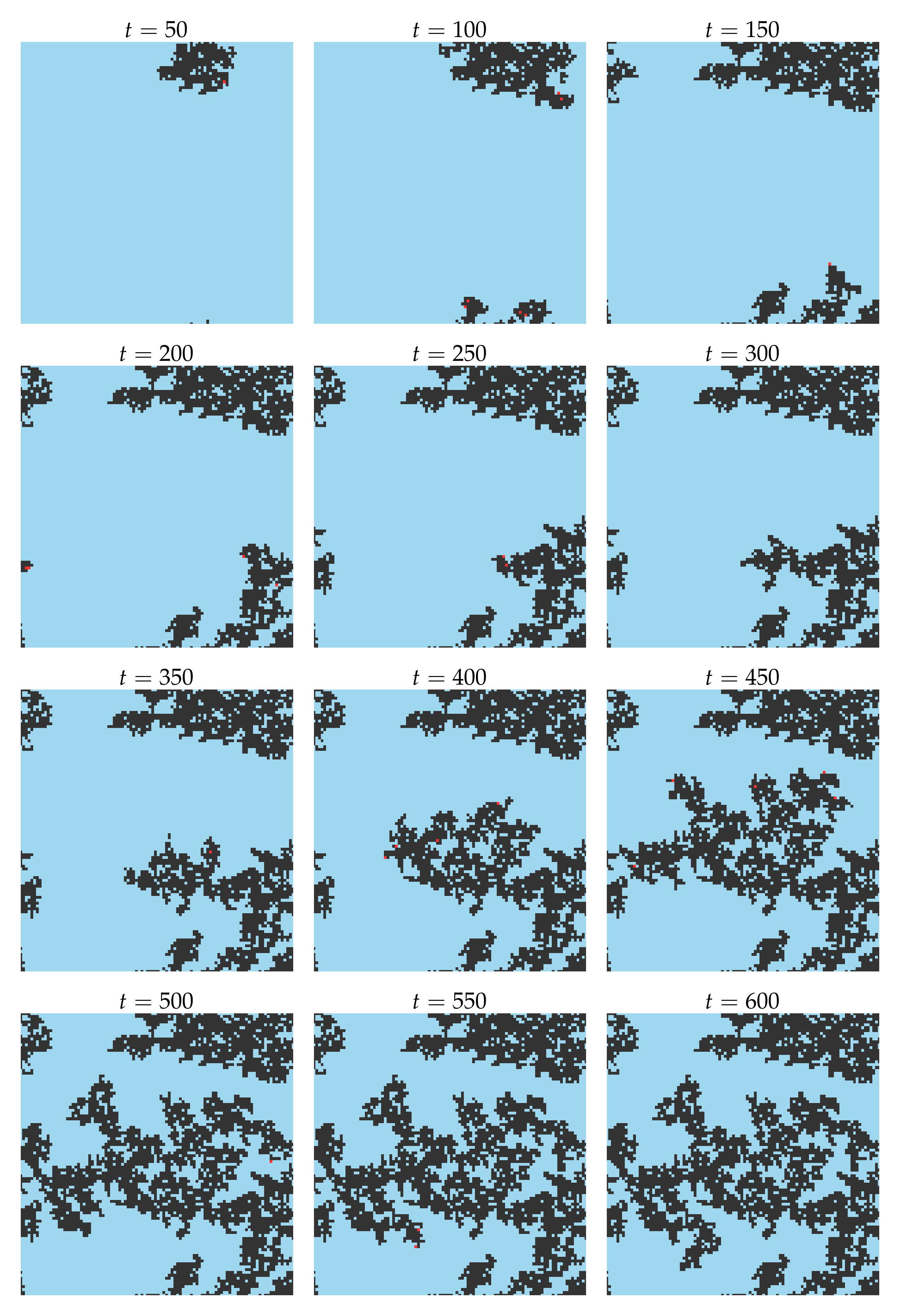}
   \caption{(Color Online) We show snapshots of one GDEP realization on a square lattice with $L=100$. The diffusive rates are $D_S = D_I = 0.5$, and the concentration is $\rho=3.682$ (close to the critical threshold $\rho_c=3.682(5)$ for this parameter set, see Tab.\ \ref{tab-thresholds}). We start from one random seed. Activated nodes are shown as black squares. Deactivated nodes are shown as sky blue and red squares, where sky blue ones do not have infected individuals inside, and red ones are for the converse. GDEP dynamics generate only one giant component, which wraps the lattice between Monte Carlo times $t=400$ and $t=450$ for this realization.}
   \label{gdep-snapshots}
\end{figure*}

Snapshots of one realization of the growing process are shown in Fig.\ \ref{gdep-snapshots}. We choose the case $D_S=D_I=0.5$ with concentration $\rho=3.682$ (close to the critical threshold $\rho=3.682(5)$, see Tab.\ \ref{tab-thresholds}). One can see that the cluster activity is concentrated at the cluster boundaries as expected. The growing process generates one cluster, which wraps the lattice between Monte Carlo times $t=400$ and $t=450$ for this particular GDEP realization. 

The system displays a continuous phase transition if the basic reproduction number $R_0 = \frac{\mu_c}{\mu_r}$ satisfies $R_0 \geq 1$. Moreover, the critical threshold $\rho_c$ increases with the relative diffusion ratio $D_S/D_I$ and vice versa. We considered in detail the cases (1) $D_S=0.25$, $D_I=0.75$, (2) $D_S=0.5$, $D_I=0.5$, and (3) $D_S=0.75$, $D_I=0.25$, with critical thresholds given in Tab.\ \ref{tab-thresholds}. The critical threshold is consistent with a minor concentration needed for a percolating phase if infected individuals can diffuse faster.
\begin{table}[ht!]
   \begin{center}
      \begin{tabularx}{\columnwidth}{Y|Y}
         \hline
         Diffusion rates                 & Epidemic threshold \\
         \hline
         $D_S = 0.25$, $D_I=0.75$ & $\rho_c = 2.922(5)$ \\
         $D_S = 0.5$ , $D_I=0.5$  & $\rho_c = 3.682(5)$ \\
         $D_S = 0.75$, $D_I=0.25$ & $\rho_c = 5.331(5)$ \\ 
         \hline         
      \end{tabularx}
   \end{center}
   \caption{The table summarizes our estimates for the critical thresholds $\rho_{c}$ on each diffusion regime: (1) $D_S<D_I$, (2) $D_S=D_I$, and (3) $D_S>D_I$. The critical threshold increases with the relative diffusion ratio $D_S/D_I$. We considered three cases, one for each diffusion regime. Finite-size results of GDEP are compatible with the dynamic percolation universality class irrespective of relative diffusion rates, as evidenced in Figs. \ref{fig-averages-cluster} and \ref{fig-averages-removed}.}
   \label{tab-thresholds}
\end{table}

The finite-size scaling behavior given in Eqs.\ (\ref{cluster-scaling}) and (\ref{removed-compartment-scaling}) should be compatible with the exact 2d percolation critical exponents $\beta = 5/36$, $\gamma = 43/18$, and $\nu=4/3$ on the square lattice. Also, the dynamic critical behavior is compatible with the 2d dynamical exponents $z=1.1309(1)$, $\delta=0.0921(3)$, and $\eta=1.5846(2)$. In the following, we show results of the asymptotic behavior $t \to \infty$ of cluster and removed concentration observables. In addition, we also show in the following the GDEP properties as a function of the time at the critical threshold $\rho_c$ that separates the percolating and non-percolating phases.

\begin{figure*}[ht!]
   \includegraphics[scale=0.35]{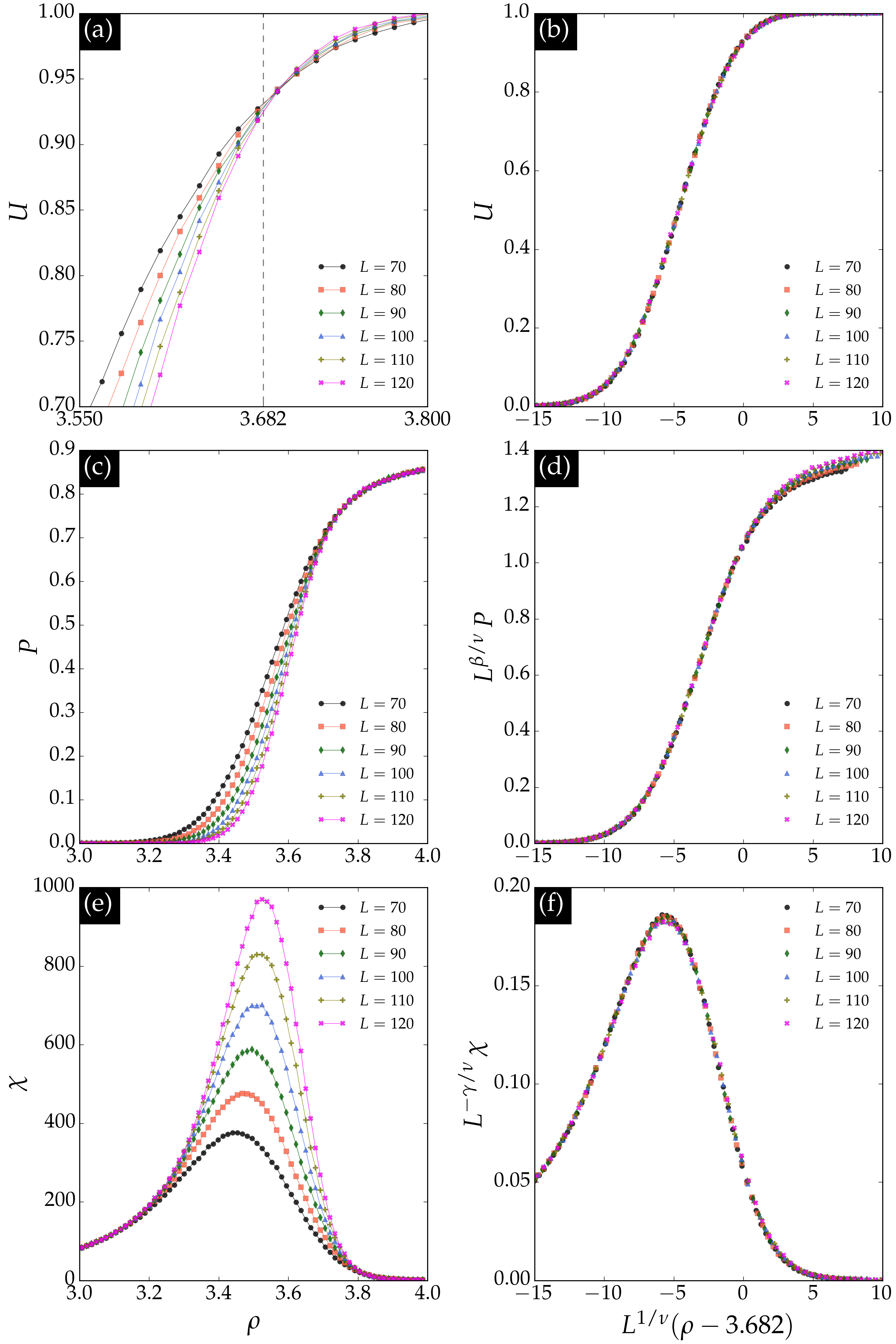}
   \caption{(Color Online) We show results for the Binder cumulant $U$ in Eq.\ (\ref{bindercumulant}), the wrapping probability $P$ in Eq.\ (\ref{orderparameter}), and the susceptibility $\chi$ in Eq.\ (\ref{susceptibility}) in panels (a), (c), and (f), respectively. Moreover, we show the data collapses of $U$, $P$, and $\chi$ according to Eq.\ (\ref{cluster-scaling}) in panels (b), (d), and (f), respectively. The data collapses are compatible with the critical threshold $\rho_c=3.682(5)$ and the 2d dynamic percolation exponent ratios $\beta/\nu=5/48$, $\gamma/\nu=43/24$, and $1/\nu = 3/4$.}
   \label{fig-averages-cluster}
\end{figure*}

We show results in the asymptotic limit $t \to \infty$ for the cluster observables in Fig.\ \ref{fig-averages-cluster} for the case (2) with $D_S = D_I = 0.5$. The system presents a continuous phase transition in the dynamic percolation universality class where the system goes from a non-percolating phase to a percolating one by increasing the concentration $\rho$. The same behavior is found for the cases (1) $D_S = 0.25$, $D_I=0.75$, and (3) $D_S = 0.75$, $D_I=0.25$ (data not shown). We can conclude that permanent immunity is the fundamental ingredient that lets the system fall into the dynamic percolation universality class. A remaining question is how the system goes from one of the standard DEP universality classes to the dynamical percolation criticality if one introduces a reinfection rate\ \cite{grassberger_spreading_1997}, which can stimulate further studies on the partial immunity of diffusive metapopulation models.

\begin{figure*}[ht!]
   \includegraphics[scale=0.35]{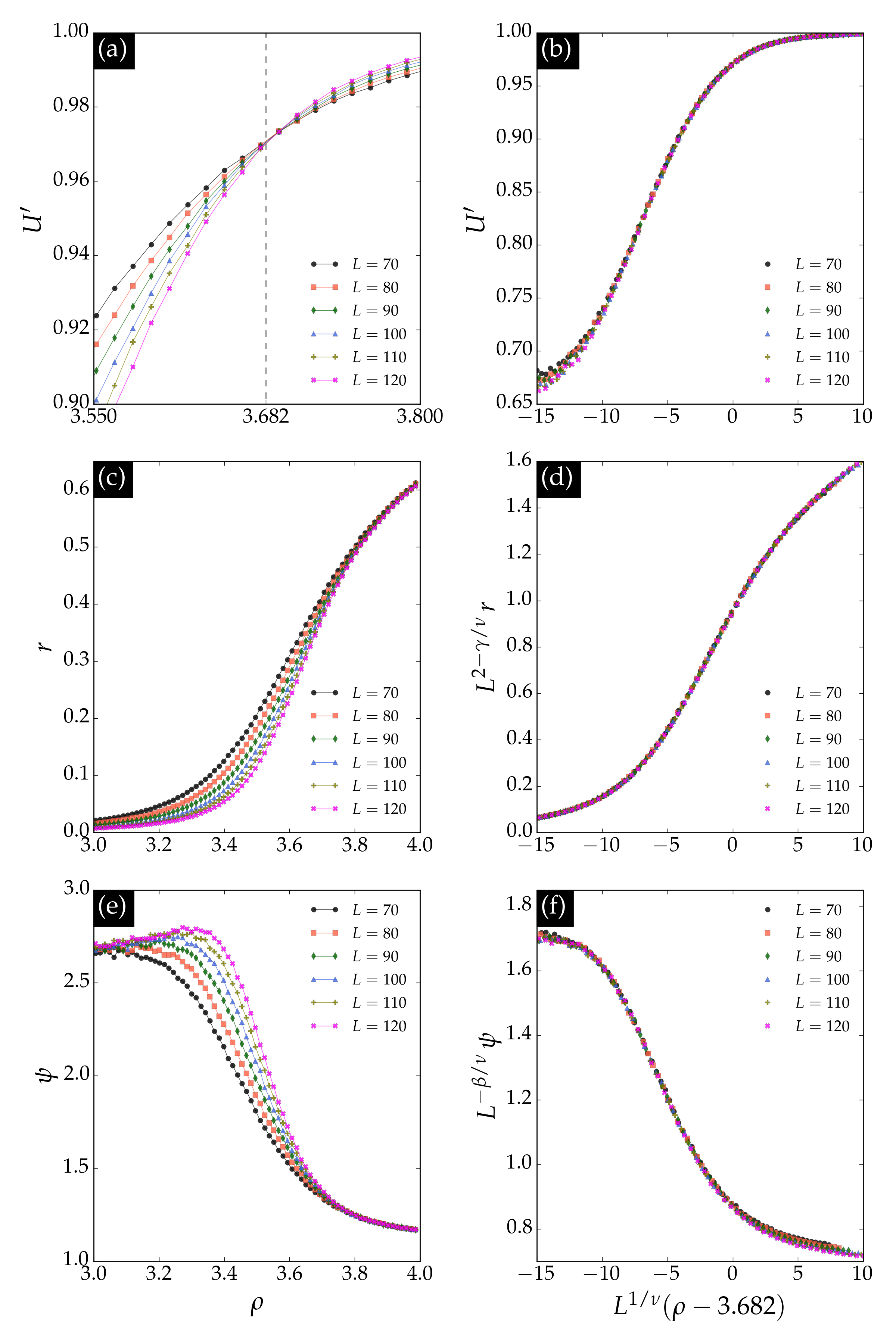}
   \caption{(Color Online) We show results for the epidemic spreading observables of Eq.\ (\ref{observables-infection}), namely the removed cumulant ratio $U^\prime$, the average removed concentration $r$, and the removed ratio $\psi$, in panels (a), (c), and (f), respectively. Moreover, we show the data collapses of $U^\prime$, $r$, and $\psi$ according to Eq.\ (\ref{removed-compartment-scaling}) in panels (b), (d), and (f), respectively. The data collapses are compatible with the critical threshold $\rho_c=3.682(5)$ and the 2d dynamic percolation exponent ratios $\beta/\nu=5/48$, $\gamma/\nu=43/24$, and $1/\nu = 3/4$.}
   \label{fig-averages-removed}
\end{figure*}

The continuous phase transition can also be analyzed by looking at the asymptotic removed concentration observables, shown in Fig.\ \ref{fig-averages-removed}. These observables can be particularly useful when the system has no proper boundary conditions to identify a wrapping cluster, which should be valid for complex and scale-free networks\ \cite{alencar_droplet_2023-2}. Of course, the critical behavior is also compatible with dynamical percolation exponents. The estimated critical concentration $\rho_c$ is the same as shown in Fig.\ \ref{fig-averages-cluster}.

\begin{figure*}[ht!]
   \includegraphics[scale=0.35]{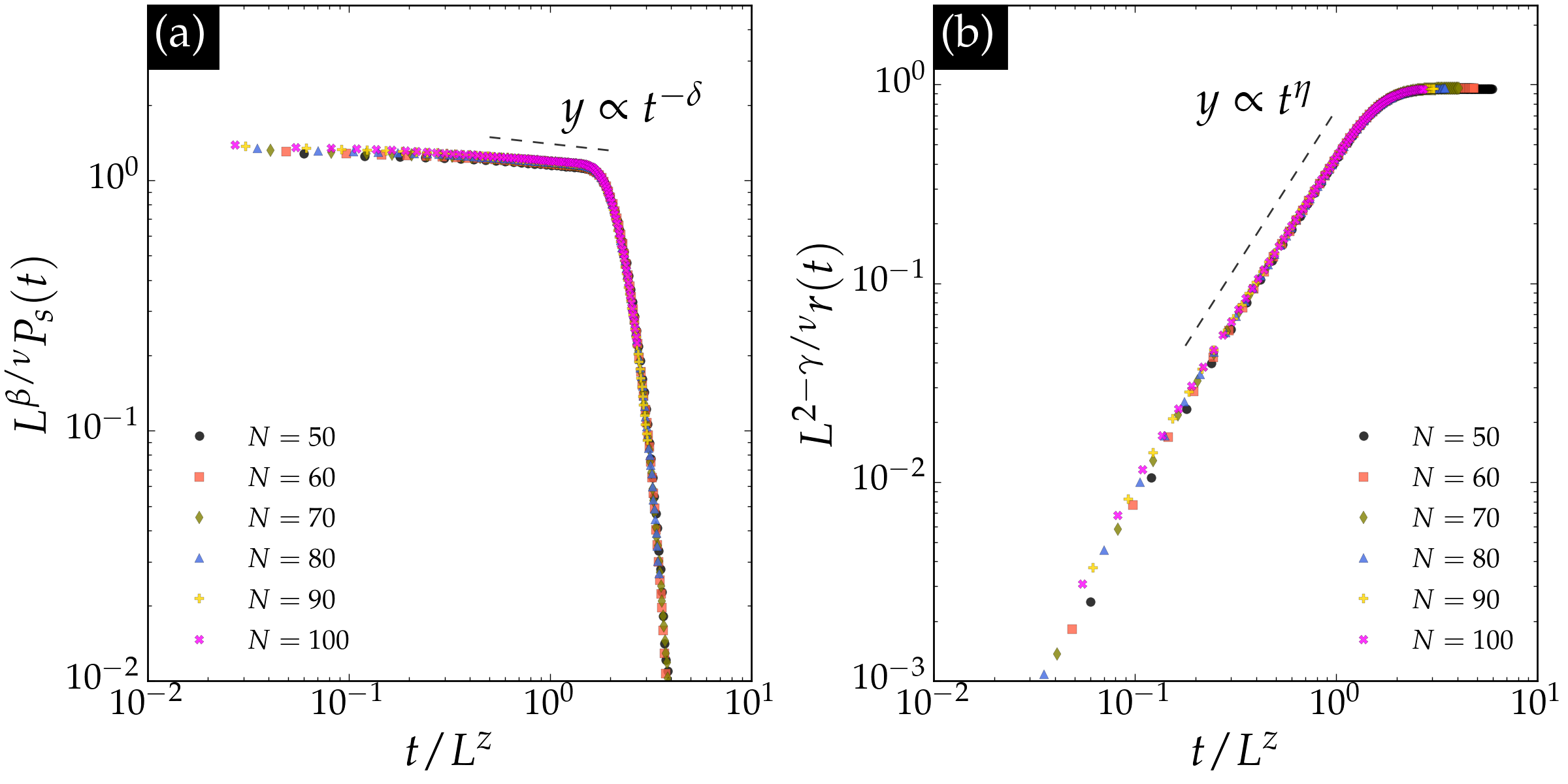}
   \caption{(Color Online) We show results for the rescaled survival probability $P_s(t)$ and the rescaled removed concentration $r(t)$ in panels (a) and (b), respectively, at the critical threshold $\rho_c=3.682(5)$. The data collapses are compatible with finite-size relations written in Eq.\ (\ref{dynamical-scaling}) and the 2d dynamic percolation exponent ratios $\beta/\nu=5/48$ and $\gamma/\nu=43/24$. The survival probability decays as a power-law with the 2d dynamical percolation exponent $\delta=0.0921(3)$ while the removed population grows as a power-law with the critical exponent $\eta=1.5846(2)$, according to Eq.\ (\ref{dynamical-properties}). Also, the time relaxation is governed by the 2d dynamical percolation exponent $z=1.1309(1)$.}
   \label{fig-dynamic-scaling}
\end{figure*}

Regarding dynamical exponents, we show results for the survival probability $P_s(t)$ and the removed concentration $r(t)$ in Fig.\ \ref{fig-dynamic-scaling}. From the scaling relations written in Eq.\ (\ref{dynamical-scaling}), we can see that the data has the expected behavior of Eq.\ (\ref{dynamical-properties}) where the survival probability decays algebraically with the exponent $\delta$ while the removed concentration grows as a power-law with the critical exponent $\eta$. The removed population at $\rho_c$ presents subexponential (power-law) growth, a feature shared by other epidemic models with short-range interactions\ \cite{paoluzzi_single-agent_2021}, which explains the flattening of the COVID-19 growing curve\ \cite{verma_covid-19_2020, maier_effective_2020} as a consequence of containment measures. 

\section{Conclusions}\label{sec-conclusions}

We considered a non-sedentary metapopulation model, namely the generalized diffusive epidemic process. This process modifies the standard diffusive epidemic, including permanent immunity. Moreover, the process does not conserve the active population. The Markovian dynamics rules were formulated as a cellular automaton with synchronous updates of all lattice nodes. The reaction-diffusion dynamics is separated into two stages: (i) the Brownian diffusion, where the particles can jump to neighboring nodes, and (ii) contamination and recovery reactions. We also map the metapopulation model as a growing process by activating nodes with successful contaminations, interpreted as infection sources.

The generalized diffusive epidemic process was studied using extensive Monte-Carlo simulations. Our results indicate a continuous phase transition in the dynamic percolation universality class controlled by the population size, irrespective of relative diffusion rates $D_S$ and $D_I$. Conversely, the standard DEP (without immunity) should fall in three different universality classes\ \cite{tarpin_nonperturbative_2017}, one for each diffusion regime. Therefore, the critical behavior of the generalized epidemic process differs from the standard diffusive epidemic process, where the universality class depends on the relative ratio of the diffusive rates. Moreover, the removed population at $\rho_c$ presents subexponential (power-law) growth related to containment measures of an epidemic spreading. 

\section{Acknowledgments} \label{sec:acknowledgements}

We would like to thank CAPES (Coordenação de Aperfeiçoamento de Pessoal de Nível Superior), CNPq (Conselho Nacional de Desenvolvimento Científico e tecnológico), and FAPEPI (Fundação de Amparo à Pesquisa do Estado do Piauí) for the financial support. We acknowledge the \textit{Dietrich Stauffer Computational Physics Lab.}, Teresina, Brazil, and \textit{Laborat\'{o}rio de F\'{\i}sica Te\'{o}rica e Modelagem Computacional - LFTMC}, Teresina, Brazil, where we performed the numerical simulations.

\bibliography{textv1}

\end{document}